# Energy-Efficient Wireless Interconnection Framework for Multichip Systems with In-package Memory Stacks

Md Shahriar Shamim, M Meraj Ahmed, Naseef Mansoor, Amlan Ganguly
Rochester Institute of Technology
Rochester, NY, USA.
(ms5604, ma9205, nxm4026,axgeec)@rit.edu

*Abstract*— **Multichip systems with memory stacks and various processing chips are at the heart of platform based designs such as servers and embedded systems. Full utilization of the benefits of these integrated multichip systems need a seamless, and scalable in-package interconnection framework. However, state-of-the-art inter-chip communication requires long wireline channels which increases energy consumption and latency while decreasing data bandwidth. Here, we propose the design of an energy-efficient, seamless wireless interconnection network for multichip systems. We demonstrate with cycle-accurate simulations that such a design reduces the energy consumption and latency while increasing the bandwidth in comparison to modern multichip integration systems.**

*Keywords— Wireless Interconnect, Multichip, Network-on-Chip*

## I. INTRODUCTION

Multichip computing modules with several chips integrated with memory banks can be found in a wide range of platform based designs from servers to embedded systems. These chips can be processing chips such as multicore chips, CPUs or GPUs or a heterogeneous mix of such chips (e.g. AMD's Fusion Accelerated Processor Units (APUs)) depending upon desired functionality. Due to scaling up of number of individual computing nodes by several orders of magnitude in these systems, the interconnection between them has become increasingly complex. Moreover, to satisfy the memory bandwidth demands, integration of in-package memory has become a norm in these systems. Integrating memory within a single package can be done either by placing memory which itself will possibly be vertically stacked on top of a multicore die i.e. monolithic 3D integration [1] or placing them side-by-side on the same substrate or interposer i.e. 2.5D integration [2]. However, in 3D stacked approach, the amount of memory that can be integrated into the package is limited by the size of multicore die (increasing the die size generally reduces yield, and hence, increases manufacturing cost). Also, the multicore processing chips need to be thinned to accommodate Through-Silicon-Vias (TSVs) through it which can induce die cracking and structural yield issues. Moreover, as the integration of the memory will essentially block the path of heat flow of the multicore die, the average die temperature of such system can become prohibitively high [3]. As a result, this approach requires sophisticated thermal management techniques. Alternatively, in horizontal or 2.5D integration, the amount of memory that can be integrated is not bounded by the size of the multicore die, rather limited by the size of substrate board or interposer. As a result, it can provide more memory capacity. Moreover, this integration technique will allow disintegration of a large multicore processing chip into several smaller processing chips. Consequently, for the same computational capabilities, disintegration will lower the total manufacturing cost considering the fact that smaller die size will eventually result in higher yield and better packing of the rectangular die on a circular wafer [3]. It also enables an easy integration of heterogeneous chips and technologies on the same platform. All these benefits of the 2.5D integration makes it a nearer term solution for a multichip system with in-package memory.

While a 2.5D integration enables integration of multiple processing chips and memory stacks, there are several challenges to the interconnection of such a system that need to be addressed. Recent trends according to the International Technology Roadmap for Semiconductors (ITRS) (http://www.itrs2.net/) predict that the pitch of the I/O interconnects in ICs is not scaling as fast as the gate lengths or pitch of on-chip interconnects. This implies a gap in density and performance of traditional I/O systems relative to on-chip interconnections. The wiring complexity of both on-chip and off-chip interconnects exacerbates the problem by posing design challenges, crosstalk, and signal integrity issues. Moreover, in the case of disintegrated processing chips, cores that were previously on the same chip are now on different processing chips. Therefore, inter-chip communication becomes extremely important and a potential bottleneck. Besides, switching between protocols is necessary if the off-chip communication protocol is different from the on-chip one. All these factors reduce the efficiency in terms of energy consumption as well as latency and bandwidth of the data transfer between communicating components such as processing cores and memory blocks in a multichip system. Therefore, we need an energy efficient, seamless, scalable interconnection network that spans across distances of a few millimeters (single chip) to several centimeters (on a multichip environment). Integrated inter and intra-chip photonic interconnections [4] is a promising solution to the off-chip interconnection challenges of traditional I/O. However, the pitch of photonic interconnects also does not scale well due to the limitations in size of silicon-photonic devices. Moreover, this technology is challenging to integrate with standard CMOS processes typically requiring a separate photonic device layer with large footprints on the chip.

Research in recent years has demonstrated that on-chip and off-chip wireless interconnects are capable of establishing radio communications within as well as between multiple chips. Wireless data communication links up to 10m in length with multi GigaHertz bandwidths in millimeter-wave (mm-wave) bands are fabricated and demonstrated [5]. Using such on-chip antennas embedded in the chip, wireless Network-on-Chip (NoC) architectures [6] are shown to improve energy efficiency and bandwidth of on-chip data communication in multicore chips [7]. In this work, we propose to use such wireless interconnects to establish a seamless communication backbone, enabling data exchange between chips in a multichip system and in-package memory. The same communication protocols used for on-chip data transfer in the intra-chip NoC will be used for off-chip data as well, eliminating the need for protocol transfer. Wireless transceivers will be deployed inside each chip and memory stack, which will be capable of establishing direct one-hop communication with other such transceivers in the system. Here, we present the design methodologies for such multichip systems with several multicore processing chips and memory stacks and demonstrate that the proposed design outperforms state-of-the-art multichip systems through system-level simulations.

## II. RELATED WORK

According to ITRS, the pitch of chip-to-chip I/O does not scale in the same proportion as on-chip global wires. Conventionally, C4 bumps coupled with in-package transmission lines are used to interconnect chips within a multichip system [8]. However, signal quality deteriorations due to microwave effects, crosstalk coupling effects, signal reflections, and frequency-dependent line losses in the transmission line limits the number of concurrent, high-speed inter-chip I/O [9]. This limits the possible off-chip bandwidth.

Various novel interconnect technologies such as vertically integrated 3D integration [1], photonic interconnects [4], inductive or capacitive coupling based interconnects [10], utilizing the available metal layers in the interposer [2], and wireless interconnects [11] are being explored to mitigate the performance issues of conventional I/O based multichip systems. In [12] wirelessly connected multichip modules are proposed for a High- Performance Computing (HPC) environment. Wireless transceivers are also proposed to be used for low latency test delivery in pre-packaged wafers [13]. However, significantly more research needs to be conducted to overcome the bottlenecks of standard multichip integration methodologies. Here we present design methodologies of a wireless multichip system with in-package memory and compare it with state-of-the-art technologies.

## III. WIRELESS INTERCONNECTION FRAMEWORK

The proposed interconnection framework is targeted for a multichip system with multicore chips and in-package memory modules. Here we describe the topology, physical layer and communication protocols of the seamless wireless interconnection framework for multichip systems.

### A. Wireless Interconnection Topology

In the proposed framework, the intra-chip interconnection topology of each multicore chip is a traditional Mesh based NoC with switches and links. Each core in the system is considered to be attached to its NoC switch. The Mesh is chosen as it is a conventional NoC topology used in several multicore based products [14] and is relatively easy to design, verify, and manufacture. To alleviate limitations of traditional inter-chip interconnects, we propose to equip some NoC switches in each multicore chip with wireless interfaces (WIs) to realize the wireless interconnects. We deploy the WIs in the multicore chips such that we achieve a wireless deployment density that avoids long multi-hop paths between the cores and the WIs. The wireless density is defined as the number of cores within each multicore chip that are serviced by a single WI. We avoid using a very high WI density such as 1WI per core, as it will increase the area overhead and potentially reduce performance due to increased contention on the shared wireless channel. Therefore, a single WI is shared by a cluster of cores in each multicore chip. The number of clusters per chip will depend on the WI density and the total number of processing cores on the chip. The WIs are deployed at one of the central switches of each cluster as shown in Fig. 1. This WI deployment strategy corresponds to the approach that achieves the optimal minimum average distance (MAD) between all switches in an intra-chip NoC [15]. This improves the connectivity of the entire multichip system by establishing direct wireless links between internal switches. Moreover, the use of WIs to interconnect multiple chips eliminates the need to layout physical channels which makes the design reusable, scalable and modular. Each memory module is considered to be a stacked DRAM mounted on-top of a base logic die, and one WI will be deployed on this logic die. This WI will be used to communicate to and from the stacked memory modules. The layers of the memory stacks are interconnected using TSVs.

### B. Physical Layer

We propose the use of on-chip embedded miniature antennas operating in the 60 GHz mm-wave band that can be fabricated within the chip to establish direct communication channels between the chips. The chosen on-chip antenna must provide the best power gain for the smallest area overhead. Several on-chip antennas, designed to operate in the mm-wave bands have been investigated in [5][7][11]. We intend the chosen antenna to be compact as well as not directional. This is because we want to communicate between antennas, which are located in different chips and potentially at different angles with respect to each-others axes. Using fabricated prototypes, a metal mm-wave zigzag antenna has been demonstrated to possess these characteristics as they are more compact compared to a linear dipole due to the zig-zag folding of the arms. In addition, such mm-wave antennas fabricated using top layer metals are CMOS process compatible making them suitable for near-term solutions to the wired interconnect problem [5]. Such mm-wave 60GHz antennas are shown to have a bandwidth of 16GHz for both intra-chip [6] and inter-chip [11] communications links through typical dielectric packaging materials. We have adopted the design of mm-wave zig-zag on-chip antennas from [11]. These antennas are also shown to not be directional and are hence able to communicate with any other WI in a similar multichip system over the shared 60GHz wireless channel.

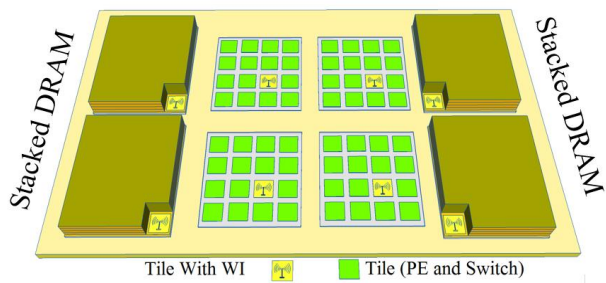

Fig. 1. Proposed wireless multichip system.

To ensure high throughput and energy efficiency, the WI transceiver circuitry must provide a very wide bandwidth as well as low power consumption. Hence, we adopt the transceiver design from [6] where low power design techniques are considered at the architecture level. Non-coherent on-off keying (OOK) modulation is chosen, as it allows relatively simple and low-power circuit implementation.

*C. Seamless Flow Control and Routing*

The routing protocol for the proposed multichip system is a seamless intra and inter-chip data communication mechanism. We adopt wormhole switching for wireline links in the multichip system where data packets are broken down into flow control units or flits [16]. All switches have bidirectional ports for all links attached to it. The WIs have an additional port equipped with the wireless transceivers to access the wireless channel. For the wireless links, we adopt the same wormhole switching with modifications explained in the next subsection.

We use a forwarding-table based routing algorithm over pre-computed shortest paths determined by Dijkstra's algorithm for both inter-chip and intra-chip data. Dijkstra's algorithm extracts a minimum spanning tree (MST) which provides the shortest path between any pair of nodes in a graph. The extracted MST depends on the start node, but the length of paths between any pair is same in all the MSTs. Hence, the MST is chosen randomly. However, for a specific start node, the shortest path along the extracted tree is always unique as the MST inherently eliminates loops. Consequently, deadlock is avoided by transferring flits along the shortest path routing tree extracted by Dijkstra's algorithm, as it is inherently free of cyclic dependencies. The route computation overheads are greatly reduced as the routing decisions are made locally based on the forwarding table only for determining the next hop and is done only for the header flit. The tail flits simply follow the reserved path as per wormhole switching.

*D. Wireless Communication Protocol*

In mm-wave interconnects wireless bandwidth is limited by the state-of-the-art transceiver design and on-chip antenna technology. To improve connectivity and performance, multiple wireless transceivers need to access the wireless medium to communicate via the energy-efficient wireless interconnects. A medium access control (MAC) mechanism enables a contention-free communication over the shared wireless channel among multiple transceivers. The authors in [7] have proposed a distributed and low-overhead token based MAC mechanism for on-chip wireless interconnects. The token based MAC grants access to the shared wireless medium to a single WI resulting in a contention free communication using the wireless channel. However, in such a MAC only whole packets are transmitted to other WIs, to maintain the integrity of the wormhole switching [11]. This increases the buffer requirement and hence static power consumption in the WIs. Therefore, we propose a MAC mechanism that allows partial packet transmission from a WI while maintaining the integrity of the wormhole switching.

In the proposed MAC, instead of circulating a token at the end of each transmission, each WI broadcasts a control packet at the beginning of its transmission. The control packet consists of a header for identification and differentiation of data packets. In addition, to enable partial packet transmission and correct routing, the control packet has 3-tuples: (*DestWI, PktID, NumFlits*) for every partial packet that it will transmit. Each 3-tuple contains the information about the number of flits (i.e., *NumFlits*) to be transmitted from the WI to a destination (i.e., *DestWI*) along with the packet ID (i.e., *PktID*) of the packet, to which the flits belong. The *PktID* enables the destination WI to identify the VC number at the destination WI to put the flits, thus maintaining wormhole switching. In case the *PktID* does not exist at the destination WI, the WI reserves an unoccupied VC. The number of 3-tuples in a control packet is limited by the number of output VC of the transmitting WI. The control packet is broadcast to all WIs. Therefore, the next WI in sequence computes the duration of the current transmission from the information in the control packet and transmits its control packet when the current transmission is completed. For this purpose, the WIs are numbered in a sequence. Thus, contention between WIs in accessing the channel is avoided.

This control packet based MAC enables an energy-efficient operation of the WIs by using sleep transistors. We adopt the design of such sleepy transceivers from [17] to put the receivers to sleep when the transmitted data is not intended for them based on the information in the control packets. This eliminates the overhead and layout complexity of the global signaling wires to carry the *sleep/$\overline{wake}$* signals as in [17].

IV. EXPERIMENTAL RESULTS

In this section, we evaluate the performance and energy efficiency of the wireless multichip interconnection systems in terms of peak achievable bandwidth per core, average packet energy, and average packet latency. The peak achievable bandwidth per core is measured as the maximum sustainable data rate in number of bits successfully routed per core per second at saturation with maximum load. Average packet energy is the energy consumed to transfer an entire packet from source to destination in the multichip system on an average. We considered the memory module to be vertically stacked 4-layered DRAM memory mounted on top of a base logic die. Each memory stack is assumed to have four channels. The base logic die works as an interface between the memory stacks and multicore chips. The delay and energy dissipation on the intra-chip wireline links is obtained through Cadence simulations considering the specific lengths of each link based on the mesh topology in each die. The wireless transceiver adopted from [6] is designed and simulated using the TSMC 65nm CMOS process and is shown to dissipate 2.3pJ/bit sustaining a data rate of 16Gbps with a signal to noise ratio (SNR) providing a bit-error rate (BER) of less than $10^{-15}$ while occupying an area of $0.3mm^2$. The switches for the proposed framework are synthesized from a RTL level design using 65nm standard cell libraries from Chip Multi Projects (http://cmp.imag.fr), using Synopsys. The delay and power dissipation including both dynamic and static power consumption of these digital components are then incorporated into a cycle accurate simulator to evaluate the performance and energy efficiency of different multichip systems. The simulator characterizes the multichip architecture and models the progress of the flits over the switches and links per cycle accounting for those flits that reach the destination as well as those that are stalled. Ten thousand iterations were performed eliminating transients in the first thousand iterations for synthetic traffic patterns separately. We also evaluate the system for real application based traffic patterns. In our experiments, we consider each core to be connected to a three-stage pipeline network switch [18]. We consider each input and output port of a switch including those with the wireless transceivers to have 8 VCs with a buffer depth of 16 flits for all the architectures considered in this paper. We consider a moderate packet size of 64 flits with a flit size of 32 bits in our experiments. All the digital components are driven by a 2.5GHz clock and 1V power supply, which are the nominal frequency and voltage in the 65nm technology node. All intra-chip wired links are considered to be single-cycle links. Next, we present the comparative evaluation of the proposed wireless interconnection framework with state-of-the-art technologies starting with a description of the architectures considered here.

*A. Architectures for Comparison*

We consider several interconnection architectures for the multichip systems for the comparative performance evaluation. We adopt the following naming convention for the architectures used in this paper: X and Y represent the number of multicore processor chips (i.e. C) and the in-package memory modules (i.e. M). In all cases, the memory stacks are considered to be mounted on both sides of the processing chip array. The architectures are described as below:

1. XCYM (Substrate): In this architecture, the processing chips and memory modules are assumed to be mounted on a substrate. The memory-chip (M-C) communications between adjacent chips occur through wide wireline memory I/Os whereas high speed serial I/Os are used for chip-chip (C-C) communications. However, between a particular pair of chips, there is only a single inter-chip link between switches at the center of the adjacent boundaries to eliminate signal crosstalk between parallel high-speed I/Os. The intra-chip communication occurs through a Mesh based NoC.

2. XCYM (Interposer): This architecture is adopted from [2] to evaluate a system with higher C-C bandwidth where the processing chips and memory modules are assumed to be placed on a silicon interposer. This interposer works as a medium to provide the point-to-point interconnects between the adjacent processing chips utilizing available metal layers in the interposer and thus extending the mesh NoC over two separate layers of silicon spanning multiple chips.

3. XCYM (Wireless): In this architecture, the processing chips and memory stacks can be mounted either on a substrate or on an interposer. The C-C and M-C communications use wireless links and is agnostic to the underlying platform.

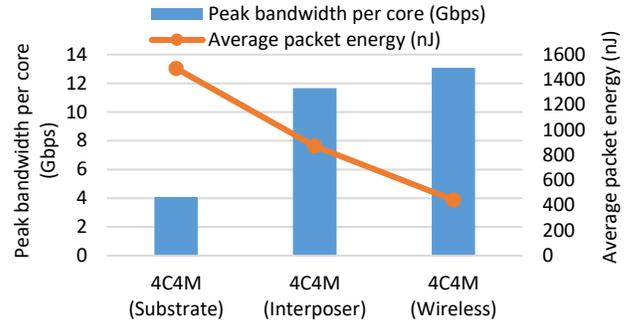

Fig. 2. Peak achievable bandwidth per core and average packet energy with uniform random traffic.

In the case of wireline configurations, the memory stacks are connected to the I/O modules of the processing chips through 128 bit (assuming µ-bump pitch of 50µm and 10mm die edge) wide channel operating at 1GHz. Hence, this wide I/O provides a total bandwidth of 128Gbps per DRAM stack with its neighboring processing chip with an energy consumption of 6.5pJ/bit [19]. The chip-to-chip high speed serial I/O channels are adopted from [8] and are shown to have a bandwidth of 15Gbps with an energy consumption of 5pJ/bit. Therefore, the energy dissipation per bit of the I/O module is higher than that of the wireless physical layer. The energy consumption of data transfer inside a memory stack is ignored as it is same in all the configurations.

*B. Performance Evaluation with Uniform Random Traffic*

In this subsection, we evaluate the performance of the multichip systems with wireless interconnections using uniform random synthetic traffic pattern at network saturation. In this case, traffic originating from each core has a certain preset probability of being a memory access while the rest of the traffic is addressed to all other cores in the entire system with equal probability. The proportion of memory access is considered to be 20% for the results presented in Fig. 2. However, the dependence of performance on this proportion is studied in the next subsection. For these evaluations, we considered a 64-core multicore system disintegrated into four 16-core processing chips connected to 4 in-package DRAM memory stacks (i.e. 4C4M). Each chip is considered to be 10mmx10mm to model the intra-chip wireline links. This makes the total processing area 400mm$^2$, which is representative of large multicore chips. This system is represented in Fig. 1. We have used wireless deployment density of 1WI per 16 cores arranged in a 4X4 array. From Fig. 2, it can be observed that the 4C4M (Wireless) have higher bandwidth per core and lower average packet energy compared to both 4C4M (Substrate) and 4C4M (Interposer). This is because in a multichip environment with uniform traffic, a significant volume of the traffic is inter-chip traffic as there are more cores collectively in other chips. As a result, a large proportion of traffic need to travel to and from the internal cores to the peripheral I/O modules and then get routed over M-C and C-C links and again travel to internal nodes at the destination processing chip. Moreover, if a core inside a multicore chip needs to communicate with a distant memory module or core, the data packets need to travel through multiple intermediate processing chips as chips are only connected to adjacent chips or memory stacks. In contrast, in the wireless multichip configuration, the WIs connect distant processing chips or memory stacks directly over energy-efficient single-hop links. This is the main architectural factor behind the gains in performance for the wireless multichip systems. The advantages of using wireless interconnection are more evident in Fig. 3 where the average packet latency for the various multichip systems with uniform random traffic for varying packet injection load is shown. Due to different average distances between cores in the different multichip interconnection architectures, the average latency characteristics are different. This is demonstrated by the average latencies at low injection loads. It can be observed that the wireless multichip has the lowest latency compared to the systems with wireline interconnections because of the shorter average path lengths due to WIs located inside the chips.

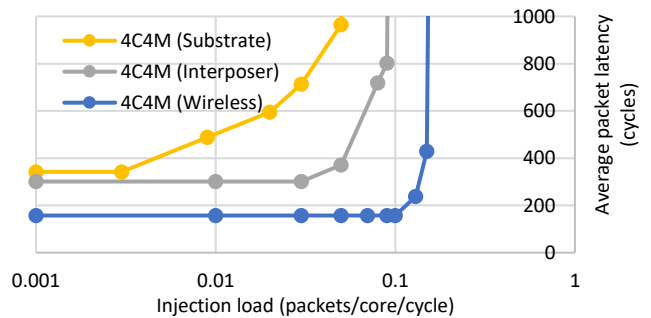

Fig. 3. Average packet latency with uniform random traffic.

The interposer based configuration has higher performance than the substrate based system. This is because the interposer based extended mesh provides a higher bandwidth by utilizing available metal layers in the interposer. Hence, we consider this configuration as a baseline wireline configuration for the following subsections.

*C. Performance Evaluation with Non-uniform Traffic*

The performance of multichip systems largely depends on their spatio-temporal characteristics such as the volume of chip-to-chip and chip-to-memory traffic. Here, we investigate the effect of variation in both chip-to-chip as well as chip-to-memory traffic on the performance of the multichip system.

The chip-to-chip traffic increases as larger multicore chips are disintegrated into small chiplets. Therefore, to evaluate the effect of the variation in chip-to-chip traffic together with disintegration, we vary the number of multicore chips keeping the total number of cores and memory modules constant. We consider three different configurations for this experiment: 1C4M (1WI/16 cores), 4C4M (1WI/16 cores) and 8C4M (1WI/8 cores) and evaluate the relative gains of the wireless multichip system with respect to the interposer based wireline system. To ensure wireless connectivity between all chips we have considered a wireless deployment density of 1WI per chip in the 8-chip system. Each memory stack is equipped with a WI for all the configurations. For this experiment, the traffic originating from each core will have a constant proportion of memory access of 20%. The inter-core (non-memory) traffic is uniformly distributed among all the other cores in the multichip system. Hence, the proportion of off-chip traffic for these configurations will be 20%, 80%, and 90% respectively. For all the configurations studied here, the combined active processing chip area is considered to be as the 4C4M configuration studied in the previous subsection. The percentage gain representing the decrease in packet energy and an increase in bandwidth of the wireless multichip system with respect to the interposer based system are shown in Fig. 4. It can be observed that the systems with wireless interconnections have lower average packet energy and higher saturation bandwidth per core compared to the interposer based wireline multichip system for all traffic scenarios. For a single-chip case, when there is only memory traffic going outside the processing chips, the configurations with wireless interconnects have better performance compared to the solely wireline I/Os, as even intra-chip traffic uses the wireless links if it reduces the path length according to the shortest path routing. This is in agreement with several mm-wave wireless intra-chip NoC papers [6][7]. However, as we increase the number of chips, the percentage gain in bandwidth and packet energy diminishes. This is because elimination of the interposer based interconnections results in loss of greater bisectional bandwidth with an increase in number of chips. On the other hand, the physical bandwidth of the wireless interconnections remains constant regardless of the number of chips. Still, we can see around 11% gain in bandwidth and 37% gain in energy efficiency compared to the interposer based wireline system with 90% off-chip traffic (8C4M).

To investigate the effect of M-C traffic, we vary the memory access keeping the system size and number of memory modules constant. To capture the variation of memory access, we vary the percentage of the traffic generated from all the cores to the memory modules. As a case study, we consider the 4C4M configuration. Fig. 5 shows the relative gain in bandwidth and packet energy compared to the interposer based wireline system as the memory traffic is varied from 20% to 80%. As we vary the memory traffic, the role of the memory-to-chip interconnections become more evident. As the interposer based system provides higher bisectional physical bandwidth between chips and memory modules, the relative performance gains of the wireless interconnection system decrease as shown in Fig. 5. This trend is similar to that with an increase in C-C traffic. However, we note that the reduction in relative gains for both bandwidth and average packet energy display an asymptotic behavior with an increase in both chip-to-chip and memory-to-chip traffic. This means that although the gains of using wireless interconnections decrease with increase in off-chip traffic, the gain will stabilize beyond a point. In our studies, the lowest gains are about 10% and 35% in peak bandwidth and average packet energy respectively.

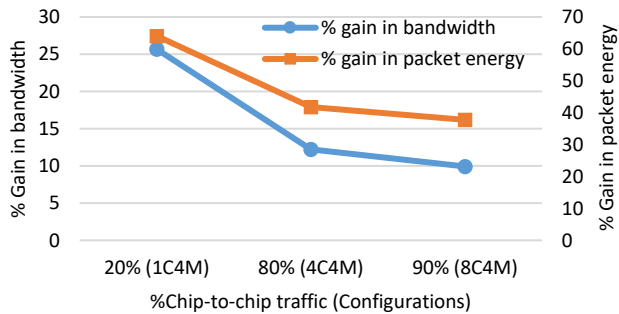

Fig. 4. Percentage gain in bandwidth and packet energy with variation in chip-to-chip traffic.

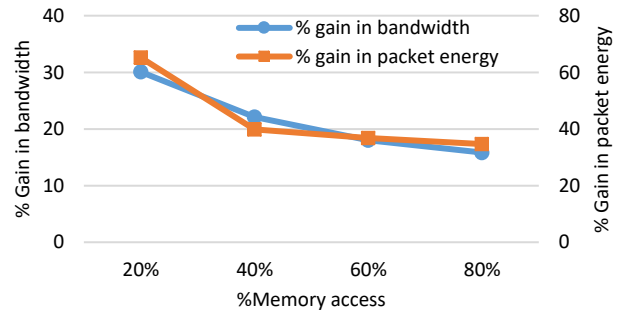

Fig. 5. Percentage gain in bandwidth and packet energy with variation in memory accesses.

*D. Performance Evaluation with Application Specific Traffic*

In this section, we evaluate the performance of the 4C4M wireless multichip configuration with application specific traffic patterns from PARSEC and SPLASH2 benchmark suites. To generate the application specific traffic patterns, we consider a multicore chip with 16 out-of-order cores with 32KB of L1 and 512KB of L2 cache running a Directory-based MOESI cache coherency protocol. These core configurations are then used to extract the core-to-memory and cache coherency traffic for these applications when they are executed till completion using SynFull [20]. In order to map these traffic patterns to the multichip environment, we consider multiple threads of the same application running on the multichip system where each processing chip executes a single thread, and the DRAM stacks are shared among threads.

The relative gain or reduction in average packet latency and average packet energy of the 4C4M wireless configuration with respect to the interposer based wireline counterpart for different application specific traffic patterns are shown in Fig. 6. The latency best represents the performance in these cases as the interconnection network is not saturated in the steady-state. The reduction in average packet latency and average packet energy for the wireless multichip system varies between applications due to the variation in off-chip traffic patterns from different memory access patterns. However, for all application-specific traffic patterns considered here, the performance of the wireless multichip system is better than the interposer based wireline configuration. The average reduction in packet latency and packet energy for the wireless multichip system is 54% and 45% compared to the interposer based system. This is due to the energy efficient single-hop wireless links connecting processing chips and memory stacks. It is worth noting that these performance benefits can be achieved with negligible active area overhead of 0.3mm$^2$ per transceiver.

## V. CONCLUSION AND FUTURE WORKS

Computing modules with multiple smaller processing chips with in-package memory stacks are becoming prevalent in platform based and HPC systems due to their performance benefits and cost-effectiveness. In this work, we explore the advantages possible if the chip-to-chip and memory-to-chip communication in multichip systems can be realized with state-of-the-art mm-wave wireless links operating in the 60GHz band. The wireless links are capable of establishing direct communication channels between cores in different chips and memory stacks via on-chip embedded transceivers and antennas. Such integration mechanism results in significant gains in performance and energy consumption in data communications in a multichip environment with negligible overhead of 0.3mm$^2$ per transceiver.

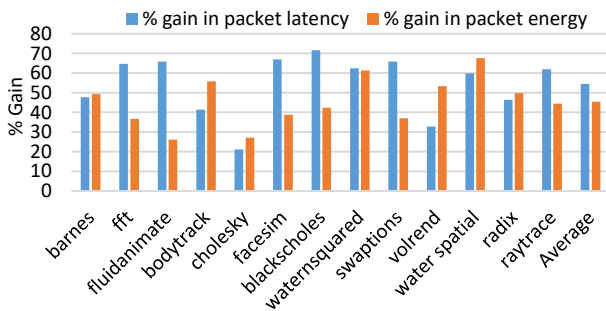

Fig. 6. Percentage gain in packet latency and packet energy with application specific traffic.